# Preferential Detachment During Human Brain Development: Age- and Sex-Specific Structural Connectivity in Diffusion Tensor Imaging (DTI) Data

Sol Lim[1,2], Cheol E. Han[1,3], Peter J. Uhlhaas[4,5,6] and Marcus Kaiser[1,2]

[1]Department of Brain & Cognitive Sciences, Seoul National University, Seoul 151–747, South Korea, [2]School of Computing Science and Institute of Neuroscience, Newcastle University, Newcastle upon Tyne NE1 7RU, UK, [3]Department of Biomedical Engineering, Korea University, Seoul 136–703, South Korea, [4]Institute of Neuroscience and Psychology, University of Glasgow, Glasgow G12 8QB, UK, [5]Department of Neurophysiology, Max-Planck Institute for Brain Research, 60438 Frankfurt a. M., Germany  and [6]Ernst Strüngmann Institute (ESI) for Neuroscience in Cooperation with Max Planck Society, Deutschordenstr. 46, Frankfurt am Main, 60528, Germany

Address correspondence to Marcus Kaiser, School of Computing Science, Claremont Tower, Newcastle University, Newcastle upon Tyne NE1 7RU, UK. Email: m.kaiser@ncl.ac.uk

Cheol E. Han, Peter Uhlhaas and Marcus Kaiser shared senior authorship

Human brain maturation is characterized by the prolonged development of structural and functional properties of large-scale networks that extends into adulthood. However, it is not clearly understood which features change and which remain stable over time. Here, we examined structural connectivity based on diffusion tensor imaging (DTI) in 121 participants between 4 and 40 years of age. DTI data were analyzed for small-world parameters, modularity, and the number of fiber tracts at the level of streamlines. First, our findings showed that the number of fiber tracts, small-world topology, and modular organization remained largely stable despite a substantial overall decrease in the number of streamlines with age. Second, this decrease mainly affected fiber tracts that had a large number of streamlines, were short, within modules and within hemispheres; such connections were affected significantly more often than would be expected given their number of occurrences in the network. Third, streamline loss occurred earlier in females than in males. In summary, our findings suggest that core properties of structural brain connectivity, such as the small-world and modular organization, remain stable during brain maturation by focusing streamline loss to specific types of fiber tracts.

**Keywords:** brain connectivity, connectome, maturation, network analysis, tractography

## Introduction

Human brain development is characterized by a protracted trajectory that extends into adulthood (Benes et al. 1994; Sowell et al. 1999; Lebel and Beaulieu 2011). Evidence from magnetic resonance imaging (MRI) has indicated a reduction in gray matter (GM) volume and thickness across large areas of the cortex and changes in subcortical structures, which may be attributed to synaptic pruning and ingrowth of white matter (WM) into the peripheral neuropil (Sowell et al. 1999, 2001; Sowell 2004; Giedd 2008; Giedd and Rapoport 2010). In contrast, WM-volume increases with age (Giedd et al. 1997, 1999; Paus et al. 1999; Bartzokis et al. 2001; Sowell 2004; Lenroot et al. 2007) which could reflect increased myelination of axonal connections (Sowell et al. 2001; Sowell 2004).

In addition to volume changes, connectivity changes of axonal fiber bundles have been investigated using diffusion tensor imaging (DTI). DTI allows the measurement of fiber integrity through estimates of fractional anisotropy (FA) and mean diffusivity (MD), which presumably relate to changes in axonal diameter, density, and myelination (Jones 2010; Jbabdi and Johansen-Berg 2011). Several studies reported increased FA and decreased MD values from childhood into adulthood in several major fiber tracts and brain regions (Faria et al. 2010; Tamnes et al. 2010; Westlye et al. 2010; Lebel and Beaulieu 2011).

Brain maturation is also accompanied by changes in the topology of structural and functional networks (Fair et al. 2009; Gong et al. 2009; Hagmann et al. 2010; Yap et al. 2011; Dennis et al. 2013). Topological features of neural networks that are now being linked to cognitive performance (Bullmore and Sporns 2009) concern their small-world and modular organization. For small-world network with brain regions or ROIs as nodes and fiber tracts as edges, there are many connections between regions mostly located nearby. At the same time, it is also easy to reach other brain regions far apart in the network due to the existence of long-range connections or shortcuts (Watts and Strogatz 1998). Therefore, small-world network shows high efficiency in facilitating information flow at both the local and the global scales (Latora and Marchiori 2001, 2003). For example, functional connectivity with high global and local efficiency correlates with higher intelligence (Li et al. 2009; van den Heuvel et al. 2009), while disrupted small-world topology is associated with impaired cognition (Stam et al. 2007; Nir et al. 2012). For a modular organization, large groups of brain regions can be considered as network modules (or clusters) if there are relatively more connections within that group than to the rest of the network (Hilgetag et al. 2000; Meunier et al. 2010). The higher connectivity within modules can segregate different types of neural information processing while fewer connections between modules allow for information integration. This community structure of the brain network incorporating and balancing both segregation and integration of neural processing has been shown to be disrupted in schizophrenia, autism and Alzheimer's disease. (Alexander-Bloch et al. 2010; de Haan et al. 2012; Shi et al. 2013).

Small-world and modular organization heavily rely on long-distance connectivity: long fiber tracts are more likely to provide shortcuts for reaching other nodes in the network and are also more likely to link different network modules (Kaiser and Hilgetag 2006). For example, connections between hemispheres or between the visual and frontolimbic network module are long distance. By providing shortcuts, long-distance connections reduce transmission delays and errors, consequently enabling synchronous and more precise information processing. Conversely, a reduction in long-distance connectivity is well known to impair cognitive ability by adversely affecting efficiency and modularity of a network (Kaiser and Hilgetag 2004). For instance, patients with Alzheimer's disease were shown to lose long-distance projections leading to an increase in functional characteristic path length (Stam et al. 2007). In addition to long-distance connections, intermodule connections, or fiber tracts linking different modules are also important to keep the community structure of brain networks and these also provide shortcuts for communicating with other functional or structural modules. Reduced between-module connectivity was strongly associated with cognitive impairment in Alzheimer's patients (de Haan et al. 2012).

Emerging data suggest that small-world topology and modular organization in brain networks are already present during early development (Fan et al. 2011; Yap et al. 2011) and that these core features of brain networks are retained during brain maturation despite significant ongoing anatomical modifications. (Bassett et al. 2008; Fair et al. 2009; Gong et al. 2009; Supekar et al. 2009; Hagmann et al. 2010). Thus, we hypothesized that certain types of fiber tracts may have been preferentially affected during development to retain important topological features during development. These potentially spared fiber tract types are likely to include long-distance connections but also fiber tracts composed of fewer streamlines and intermodule fiber tracts. Fiber tracts of the latter 2 types are often, but not necessarily, also long-distance connections (Supplementary Material S5 and Fig. S4). Therefore, we analyzed all 3 types of fiber tracts in relation to topological changes.

To test our hypothesis, we obtained DTI data from a large cohort of subjects between 4 and 40 years and constructed streamlines from deterministic tractography to identify fiber tracts in cortical and subcortical networks. Our results show that the number of streamlines decreased overall with age while small-world and modular parameters did not change. Specifically, our results showed that streamline loss occurred mostly for fiber tracts composed of more than average number of streamlines, short and within-module/within-hemisphere fiber tracts. This focus on certain types of fiber tracts goes beyond what would be expected by a type's prevalence within a network suggesting a preferential detachment of streamlines. In addition to modifications in cortical fiber tracts, pronounced changes were observed in subcortical structures, such as basal ganglia and anterior cingulate cortex (ACC). Finally, streamline reductions occurred at an earlier age in females than in males, suggesting sex-specific maturation of connectivity patterns during human brain maturation.

## Materials and Methods

### DTI Data
We made use of a public DTI database (http://fcon_1000.projects.nitrc.org/indi/pro/nki.html) provided by the Nathan Kline Institute (NKI) (Nooner et al. 2012). DTI data were obtained with a 3 Tesla scanner (Siemens MAGNETOM TrioTim syngo, Erlangen, Germany). $T_1$-weighted MRI data were obtained with 1 mm isovoxel, FoV 256 mm, TR = 2500 ms, and TE = 3.5 ms. DTI data were recorded with 2 mm isovoxel, FOV = 256 mm, TR = 10 000 ms, TE = 91 ms, and 64 diffusion directions with $b$-factor of 1000 s mm$^{-2}$ and 12 $b$0 images. We included 121 participants between 4 and 40 years.

### Data Pre-Processing and Network Construction
We used Freesurfer to obtain surface meshes of the boundary between GM and WM from $T_1$ anatomical brain images (http://surfer.nmr.mgh.harvard.edu) (Fig. 1). After registering surface meshes into the DTI space, we generated volume regions of interest (ROIs) based on GM voxels. Freesurfer provides parcellation of 34 anatomical regions of cortices based on the Deskian atlas (Fischl et al. 2004; Desikan et al. 2006) and 7 subcortical regions (Nucleus accumbens, Amygdala, Caudate, Hippocampus, Pallidum, Putamen, and Thalamus) (Fischl et al. 2002, 2004) for each hemisphere, thus leading to 82 ROIs in total (See Supplementary Table S5 for full and abbreviated names of ROIs).

To obtain streamline tractography from eddy current-corrected diffusion tensor images (FSL, http://www.fmrib.ox.ac.uk/fsl/), we used the fiber assignment by continuous tracking (FACT) algorithm (Mori and Barker 1999) with 35° of angle threshold through Diffusion toolkit along with TrackVis (Wang et al. 2007) (Fig. 1). This program generated the tractography from the center of all voxels (seed voxels) in GM/WM except ventricles; a single streamline started from the center of each voxel. Thus, the number of total streamlines never exceeds the number of seed voxels.

In addition, we also performed tractography with the following parameters: a single tracking per voxel for 45° threshold and 10 random trackings per voxel for both 35° and 45° thresholds, in total 3 more cases. These additional analyses were performed to assure that the results were consistent despite varied tracking parameters (Supplementary Material S6 and Fig. S5).

For network reconstruction, we used the UCLA Multimodal Connectivity Package (UMCP, http://ccn.ucla.edu/wiki/index.php) to obtain connectivity matrices from the defined and registered ROIs and tractography, counting the number of streamlines between all pairs of defined ROIs. The resulting matrix contains the streamline count between all pairs of ROIs as its weight. We also computed the average connection lengths between ROIs (if there is no connection between a pair, the length was set to zero). The connection length of a streamline was based on its 3D trajectory.

### Network Analysis
Short explanations of network measures are provided here (for details, cf. Supplementary Material S2). Edge density represents the proportion of existing connections out of the total number of potential connections (Kaiser 2011). Note that the weights of individual edges (streamline count) might change but edge density will remain the same as long as the total number of edges (fiber tracts) is unchanged. Small-world topology can be characterized by high global and local efficiency (Latora and Marchiori 2001, 2003). Global efficiency represents how efficiently neural activity or information is transferred between any brain regions on average and local efficiency indicates how well neighbors of a region, or nodes that are directly connected to that region, are interconnected. Efficiency is greatly affected by the sparsity of the network (Kaiser 2011); when there are fewer edges and also even fewer streamlines, efficiency decreases. Thus, we normalized efficiency with values obtained by 100 randomly rewired networks where randomly selected edges were exchanged while preserving both degree and strength of each node (Rubinov and Sporns 2011). Modularity $Q$ represents how modular the network is; higher values of $Q$ indicate that modules are more segregated with fewer connections between modules. In contrast, lower $Q$ values indicate more connections between modules and thus represent more distributed organization (Newman 2006). We also compared the modular membership assignment using the normalized mutual information (NMI) (Alexander-Bloch, Lambiotte, et al. 2012). Within-module strength and participation coefficient show local changes in modular organization. Within-module strength indicates the degree to which a node is connected to others nodes in the same module (Guimera and Amaral 2005); high within-module strength implies that the node is more connected to nodes within the module in which it participates than the average connectivity of the other nodes in the module. The participation coefficient indicates how well the node is connected to all other modules with higher values if many connections of the node are distributed to other modules. We used Matlab routines from the Brain Connectivity Toolbox (Rubinov and Sporns 2010).

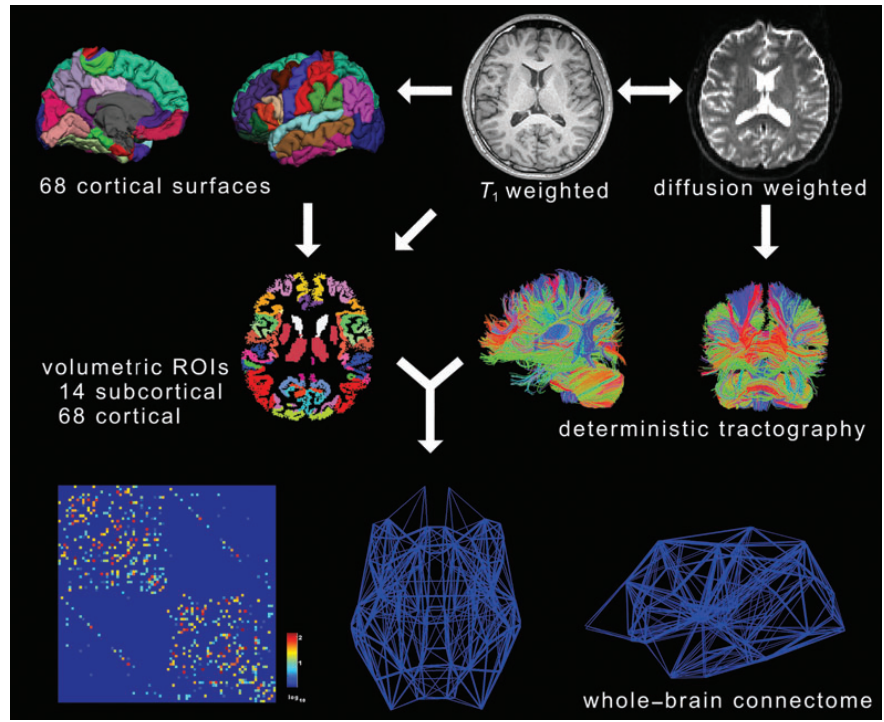

**Figure 1.** Overall procedure. From $T_1$-weighted images, we generated 82 regions of interests (ROIs, 34 cortical areas and 7 subcortical areas per hemisphere, on the left). From diffusion tensor images (DTI), we reconstructed streamlines using deterministic tracking (on the right). Combining 2 preprocessing steps, we constructed weighted networks, where the number of streamlines between any pair of ROIs formed the weight of an edge (fiber tract).

### Edge Group Analysis

We grouped fiber tracts into categories in terms of (a) the number of streamlines- (thin vs. thick), (b) the length of the streamline trajectory- (short vs. long), and (c) whether they were within modules (intramodule) or between modules (intermodule) and counted the streamlines in each group. Then, we examined with general linear model (GLM) if the number of streamlines in each category changed over age (see Statistical Analysis).

As the spatial (b) and topological (c) properties often overlap but do not always coincide (Supplementary Material S5 and Fig. S4), we investigated all 3 cases (da Fontoura Costa et al. 2007; Meunier et al. 2010). In general, short-length and intramodule edges are more numerous than others. Therefore, larger changes in those edges would occur for random selection. Accordingly, we used $\chi^2$ tests to verify any preferential detachment that goes beyond the streamline loss that would be expected based on the number of fiber tracts of each type. We standardized weights and lengths for each individual and categorized edge into 2 groups by the mean of each participant to account for differences in brain volume and size. For instance, an edge or a fiber tract for a participant is classified as "thin" when the weight of the fiber tract is less than the average weight of the participant. Likewise, a fiber tract is considered thick when the weight is above the average of the participant. The same procedure was performed to differentiate short and long fiber tracts. Therefore, types of fiber tracts were distinguished using a subject-specific threshold.

### Individual Edge Analysis

In addition to analyzing types of fiber tracts differences, we also examined changes for individual edges that included the subset of total fiber tracts that all participants had in common (128 edges, ~32.3% of the total number of edges 396 ± 20). Note that the total number of edges was around 400, which is 12% of the total number of possible connections ($n = 3321$). This proportion is consistent with previous evidence suggesting that the human brain has a sparse connectivity ranging between 10% and 15% (Kaiser 2011). To analyze individual edges, each edge with significant age-related changes was mapped to the corresponding lobe according to Freesurfer Lobe Mapping (Table 2 and Supplementary Table S4) (http://surfer.nmr.mgh.harvard.edu/fswiki/CorticalParcellation).

### Statistical Analysis

To assess how theoretical graph measures changed during development, we used GLM approach (see eqs. 1, 2, and 3). Linear and quadratic effects of age and the interaction between age and gender were investigated. The quadratic term of age, gender factor, and the interaction term between age and gender were dropped and refitted when the effects were not significant following an $F$-test as all tested models were nested. Akaike Information Criterion (AIC) and Bayesian Information Criterion (BIC) were also used for model comparison and selecting variables when the $F$-test alone did not provide a strong preference for a model. As AIC tends to prefer more complex models with a larger number of variables compared to BIC (Kadane and Lazar 2004), AIC and the $F$-test provided consistent results in general. When the results of the 3 tests conflicted, we chose the most conservative model with a smaller number of variables. Two-tailed tests were used for all analyses and tests were regarded as significant with an $\alpha$ level of 0.05. Quadratic age effect was found to be significant in a few fiber tracts but occurred less frequently than linear cases. We therefore chose to report age effects of the numbers of streamlines where decrease and increase could follow a linear or, less often, a nonlinear pattern.

$$y = \beta_0 + \beta_1 \cdot \text{age} + \beta_2 \cdot \text{sex} + \epsilon \quad (1)$$

$$y = \beta_0 + \beta_1 \cdot \text{age} + \beta_2 \cdot \text{sex} + \beta_3 \cdot \text{age} \cdot \text{sex} + \epsilon \quad (2)$$

$$y = \beta_0 + \beta_1 \cdot \text{age} + \beta_2 \cdot \text{sex} + \beta_3 \cdot \text{age}^2 + \epsilon \quad (3)$$

where $y$ is measurement, $\beta_0$ intercept (bias), $\beta_1$ slope over age, $\beta_2$ coefficient for sex difference, $\beta_3$ coefficient for interaction effect of age and sex or quadratic age effect, and $\epsilon$ represents errors (noise), which are independent and identically distributed, having a Gaussian (i.e., normal) distribution with mean zero and variance $\sigma^2$.

Through the group analysis of edges (see Edge Group Analysis), we identified which types of edges were undergoing developmental changes. Using repeated-measures GLM, we tested whether 2 groups had different slopes and $\chi^2$ tests were used for verifying the slope difference of GLM considering the proportion of each group with each individual network. For individual edge analysis, $\chi^2$ tests, and nodal properties such as within-module strength and participation coefficients, false discovery rate (FDR) procedure was used with a $q$ level of 0.05, adjusting significance level and confidence intervals (Benjamini and Hochberg 1995; Benjamini et al. 2005; Jung et al. 2011). All statistical tests were calculated in Matlab R2012b (Mathworks, Inc., Natick, MA, USA) and R (R Development Core Team 2011) with R packages (Lemon 2006; Bengtsson 2013; Sarkar 2008; Weisberg and Fox 2011; Suter 2011).

## Results

We performed a combined analysis of fiber tracts with network parameters to examine on-going changes in fiber tracts in terms of small-world topology and modularity, which may account for a relationship between topological changes and modifications in fiber tracts.

We compared developmental changes examining the following features: 1) Overall connectedness: total number of streamlines, edge density, and thin versus thick connectivity, 2) small-world organization: efficiency and short- versus long-distance connectivity, 3) modular organization: modularity and within versus between module connectivity, and 4) local organization: individual edge analysis.

### Age Effect for Both Genders

#### Connectedness

*Streamline count versus edge density.* The total number of streamlines decreased ($\beta_1 = -68.87$, $t_{(118)} = -5.796$, $P < 0.001$, Fig. 2A) with age; however, edge density remained stable ($t_{(118)} = 0.757$, $P = 0.451$, Fig. 2B).

*Thick versus thin edges.* Edge density or the number of fiber tracts could be maintained either through new fiber tracts that make up for lost fiber tracts due to streamline reduction or through sparing thin edges and therefore retaining existing fiber tracts while changing only weights for fiber tracts. To test the latter hypothesis, we tested whether there were differences in developmental patterns of thick or thin edges (see Edge Group Analysis).

Streamlines in both thick and thin edges decreased with age [thick edges: $\beta_1 = -60.184$, $t_{(118)} = -6.195$, $P < 0.001$, Fig. 2C; thin edges: $\beta_1 = -8.685$, $t_{(118)} = -3.27$, $P = 0.001$]. However, the slopes between thick and thin edges were significantly different (repeated-measures GLM, $F_{1,119} = 40.196$, $P < 10^{-8}$, Fig. 2C) with the slope of thick edges showing an ~8 times steeper slope than thin edges. This preferential reduction of streamlines within thick edges could not be explained by the frequencies of thin and thick fiber tracts ($\chi^2$ test, $P < 10^{-20}$).

### Small-World Topology and Long-Distance Connectivity

*Efficiency and small-world topology.* Global and local efficiency decreased during development (global: $\beta_1 = -0.001$, $t_{(118)} = -2.496$, $P = 0.014$, Fig. 2D, local: $\beta_1 = -0.019$, $t_{(118)} = -4.435$, $P < 0.001$, Fig. 2E). Although global and local efficiencies may have been slightly compromised by the loss of streamlines, small-world features were maintained; global efficiency paralleled that of the rewired network ($0.88 \pm 0.036$, ~0.9), while local efficiency was much higher ($4.06 \pm 0.446$, ~4) than that of the random networks.

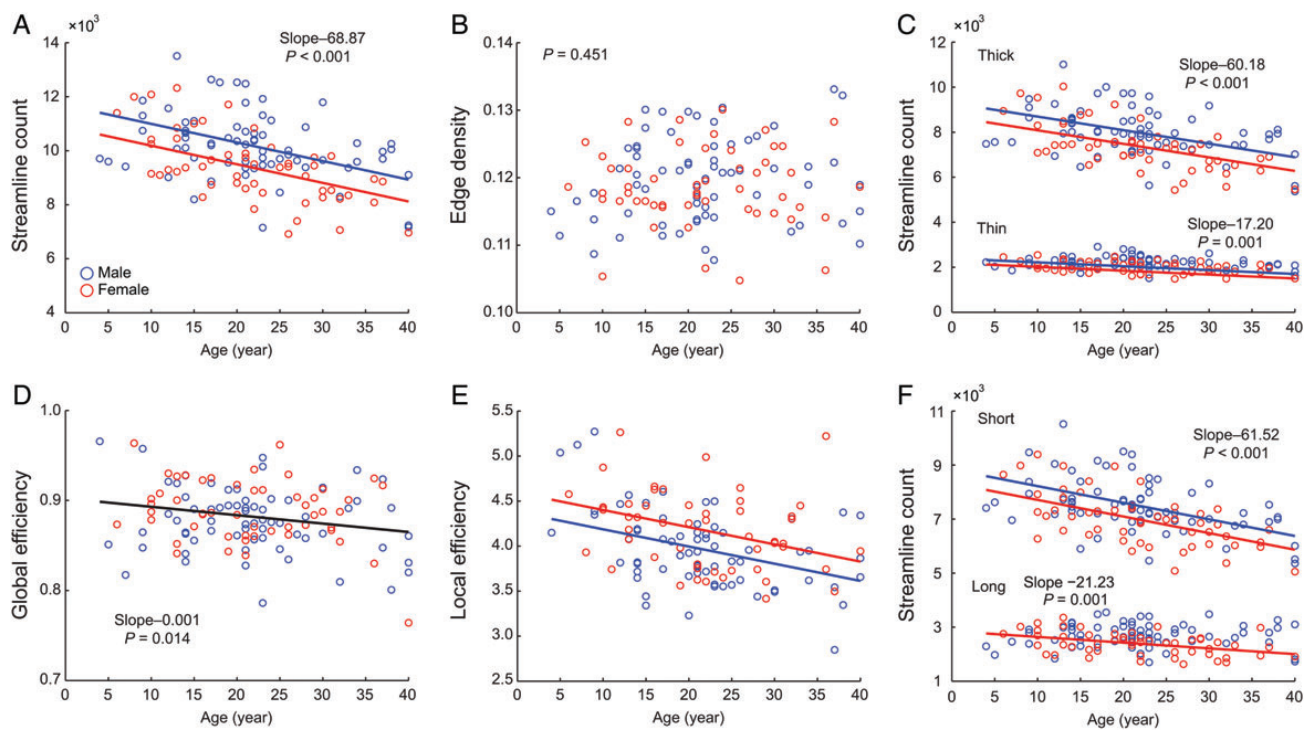

**Figure 2.** Topological and spatial network properties. Fitted lines were drawn when there was a significant age effect (red: female, blue: male). When multiple lines were drawn, the lines are parallel unless otherwise noted. Black line represents significant age affect without a sex difference. (*A*) Total number of streamlines, (*B*) edge density, (*C*) streamline count in thick versus thin edges, (*E*) global efficiency, (*F*) local efficiency, and (*G*) streamline count in short versus long streamlines.

*Short- versus long-distance connectivity.* As topological and spatial organizations are often linked (Kaiser and Hilgetag 2006; da Fontoura Costa et al. 2007; Meunier et al. 2010), we tested whether the pattern of changes in short- and long-distance connectivity corresponded to changes in efficiency. From the preserved small-world topology, we would expect long fiber tracts were likely to be conserved.

Decreasing slopes of the streamline count between short and long edges were significantly different ($F_{1, 119} = 44.965$, $P < 10^{-9}$) with short-distance connections showing a pronounced reduction (short: $\beta_1 = -61.515$, $t_{(118)} = -6.773$, $P < 10^{-9}$, Fig. 2F), which was not solely explained by a higher proportion of short-distance edges ($\chi^2$ test, $P < 10^{-6}$).

*Modular Organization*

*Modularity and module membership assignment.* Modularity did not change with age ($t_{(118)} = -1.335$, $P = 0.184$, Fig. 3A) and community structure remained stable during development (Supplementary Table S2 and Fig. S1). Overall modular organization based on the NMI did not differ across age ($P = 0.355$), and there were no significant nodal changes in membership assignment after multiple comparison correction using FDR (detailed information cf. Supplementary Material S3).

*Within-module strength and participation coefficient.* Twenty of 82 ROIs (24.4%) showed significant changes in within-module strengths and participation coefficients (FDR corrected). Overall changes were asymmetric between hemispheres, affecting homologous ROIs either in the left or right hemisphere. Ten of the 24 ROIs (42%) characterized by age effects were areas in subcortical regions, such as the basal ganglia, thalamus, and nucleus accumbens (Table 1). Specifically, within-module strengths decreased while participation coefficients increased, indicating that with development connections involving basal ganglia decreased within its module while connections to the surrounding modules/regions decreased. In contrast, 8 ROIs within the ACC and the paralimbic division (Mesulam 2000) were mainly characterized by increased within-module connectivity with age.

*Within versus between module analysis.* Modular membership and modularity Q stayed relatively stable during development although there were some ROIs that showed significant changes in terms of inter- versus intramodules connectivity (see Within-Module Strength and Participation Coefficient, Table 1). This can be realized when changes occurred mainly within modules. The decreasing slopes of streamline count for intra- and intermodule edges differed (repeated-measures GLM, $F_{1,119} = 33.186$, $P < 10^{-7}$). The reduction of streamlines occurred within modules ($\beta_1 = -61.25$, $t_{(118)} = -6.321$, $P < 10^{-8}$, Fig. 3B) but not between modules ($t_{(118)} = -1.831$, $P = 0.0696$, Fig. 3B). This preference was not fully explained by the higher proportion of intramodule edges ($\chi^2$ test, $P < 10^{-6}$).

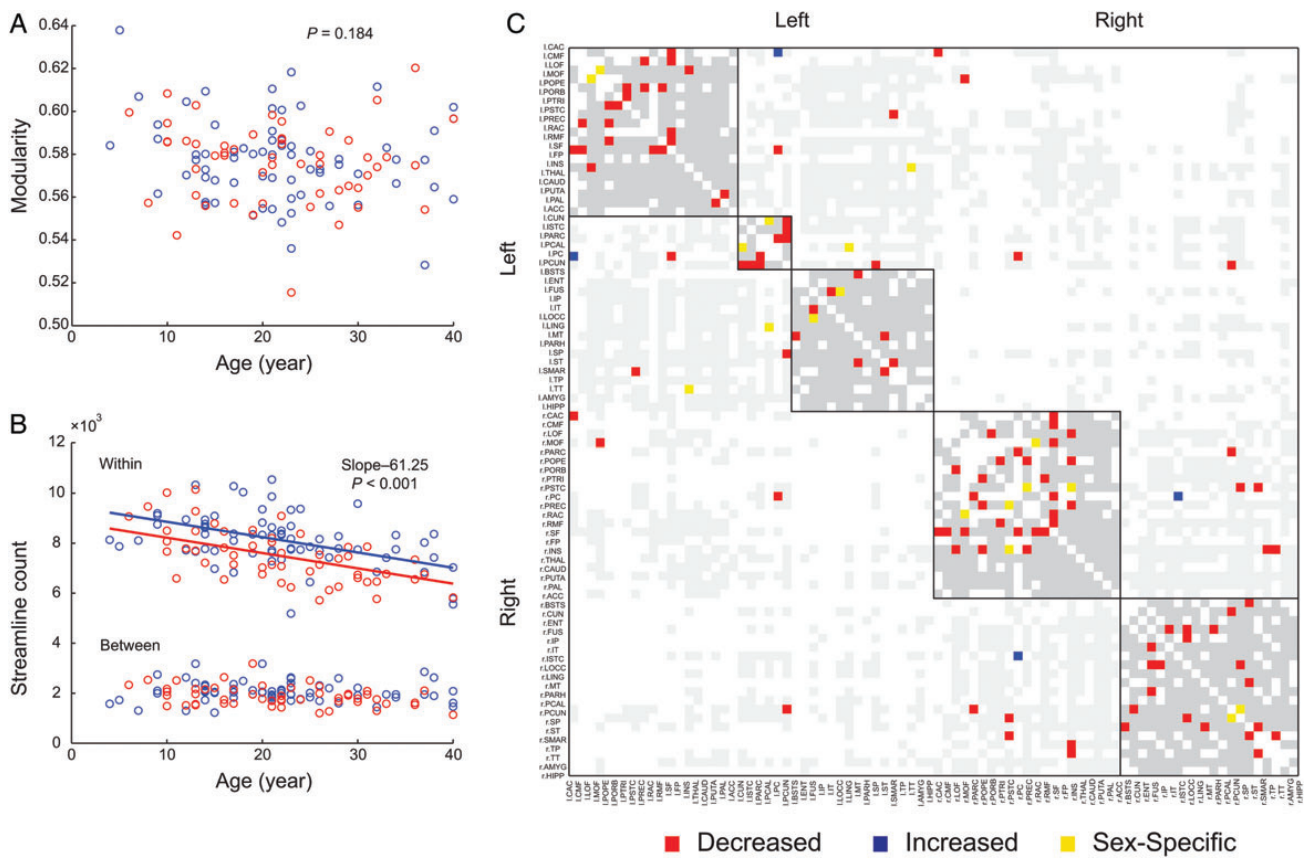

**Figure 3.** Modular organization. (A) Modularity Q, (B) Streamline count in within- versus between-module edges, and (C) individual edge analysis (gray: intramodule edges and light gray: intermodule edges, both without changes over age; red: edges with a decreased streamline count, blue: edges with an increased streamline count; and yellow: edges with sex-specific changes). When multiple lines were drawn, the lines are parallel unless otherwise noted. A list of all changes is provided in Table 2 for sex-specific changes and Supplementary Table S4 for age effect.

**Table 1**
ROIs with age effect in within-module strength (WMS) and participation coefficient (PC)

|     | Increased | Decreased | Sex-specific |
| --- | --- | --- | --- |
| WMS | **lh.caudalanteriorcingulate (F)**<br>lh.entorhinal (T)<br>lh.parahippocampal (T)<br>**rh.caudalanteriorcingulate (F)**<br>**rh.rostralanteriorcingulate (F)** | **lh.thalamus**<br>**lh.accumbens**<br>rh.putamen (f > m)<br>**rh.pallidum** | **lh.putamen**<br>M: decreased<br>rh.paracentral (F)<br>M:Increased |
| PC  | lh.putamen<br>lh.pallidum<br>rh.caudate<br>rh.putamen (m > f)<br>rh.pallidum (m > f) | **rh.caudalanteriorcingulate (F)**<br>rh.paracentral (F)<br>rh.posteriorcingulate (P) | lh.medialorbitofrontal (F)<br>M: increased<br>rh.insula (m > f) |

Note: Basal ganglia showing a more distributed network and anterior cingulate cortex showed a more focused connectivity within its module (bold).
FDR corrected, with a $q$ level of 0.05.
F, frontal lobe; P, parietal lobe; T, temporal lobe; O, occipital lobe; lh, left hemisphere; rh, right hemisphere; f, female; m, male.

**Table 2**
Edges with sex-specific age-related changes

| ROI (node) | Lobe | ROI (node) | Lobe | Sex | Slope | FDR-adjusted $P$ |
| --- | --- | --- | --- | --- | --- | --- |
| lh.cuneus | O | lh.pericalcarine | O | Male | 1.035 | 0.0002 |
| lh.fusiform | T | lh.lateralocciptal | O | Female | −0.9 | 0.041 |
| lh.lingual | O | lh.pericalcarine | O | Female | −0.535 | 0.041 |
| lh.transversetemporal | T | lh.insula | | Male | −0.908 | 0.0002 |
| rh.postcentral | P | rh.insula | | Male | −0.747 | 0.001 |
| rh.medialorbitofrontal | F | rh.rostralanteriorcingulate | P | Female | −0.769 | 0.0003 |
| rh.precuneus | P | rh.superiorparietal | P | Female | −1.351 | 0.023 |
| F:1 | P:4 | T:2 | O:5 | | | |

Note: Sex-specific developmental changes were asymmetrical compared to the developmental changes for both genders.
The last row gives an overview of how often different lobes participate in these changes. $P$ values were adjusted by FDR with a $q$ level of 0.05.
lh, left hemisphere; rh, right hemisphere; m, male; f, female; F, frontal lobe; P, parietal lobe; T, temporal lobe; O, occipital lobe.

### Individual Edge Analysis

To identify edge-specific age effects, we investigated 128 edges found in all participants (total number of edges: 396 ± 20), of which 64 edges showed significant age-related changes. The findings were consistent across different tractography parameters (Supplementary Material S6 and Fig. S5). First, 57 edges (89%) showed developmental changes: 55 edges (86%) showed a reduced number of streamlines while only 2 (3%) had an increased streamline count (Figs 3C and 5A, Supplementary Table S4). Reduction of streamlines was most pronounced in the frontal lobe; increased number of streamlines only occurred for 2 connections (3%) of cingulate cortex. These changes for both genders mainly occurred in the frontal and parietal lobe.

### Sex-Specific Age-Related Changes

Unlike developmental changes for both males and females, only several network properties showed sex-specific developmental changes. While both male and females lost short streamlines, only female participants were characterized by a decrease in long streamlines. However, this decrease was less pronounced than the reduction in short streamlines ($\beta_1 = -21.229$, $t_{(50)} = -3.372$, $P = 0.001$, Fig. 2F). While global modular organization (see Modularity and Module Membership Assignment) did not show sex differences, 3 regions of 20 showed sex-specific developmental changes in within-module strength and participation coefficients (Table 1). In the individual fiber tract analysis, changes that only affected one gender occurred in 7 fiber tracts (11%) (Figs 4–6B, Table 2). There were 4 edges with age effect only in females, and 3 edges only in males, mostly involving occipital and parietal regions.

### Sex Differences Independent of Age

Males had ~800 more streamlines than females across age ($t_{(118)} = -3.949$, $P < 0.001$, Fig. 2A) mainly due to larger brain size. In particular, males had larger number of streamlines for within-module edges (Supplementary Fig. S2). Although males showed a substantially larger number of streamlines, male and female participants demonstrated comparable edge density ($t_{(118)} = -0.880$, $P = 0.381$, Fig. 2B) as well as global efficiency (Global: $t_{(118)} = 1.598$, $P = 0.113$, Fig. 2D). However, females showed higher local efficiency than males (Local: $t_{(118)} = 2.891$, $P = 0.005$, Fig. 2E). Modularity ($t_{(118)} = -0.409$, $P = 0.684$, Fig. 3A) and overall modular organization based on the NMI also did not differ between genders ($P = 0.177$). Most ROIs did not show gender differences in within-module strength and participation coefficient except 4 ROIs (Table 1).

### Discussion

In this study, we investigated changes in structural connectivity (SC) between ages of 4–40 years from DTI data in cortical and subcortical regions. Previous studies had shown that the human brain undergoes vast structural changes involving alterations in the topology of structural and functional connectivity. Yet, core properties such as small-world topology and modular organization were retained throughout development (Fair et al. 2009; Gong et al. 2009; Supekar et al. 2009; Hagmann 2010; Dennis et al. 2013). Therefore, we examined if specific types of fiber tracts were preferentially affected, which might be conducive to conserving major topological features. Our results show that small-world features, the number of fiber tracts, and the modular organization remained largely stable over age despite a significant reduction of streamlines in fiber tracts. This reduction preferentially affected fiber tracts that were relatively short, consisted of more streamlines and were within topological modules (Fig. 7*A*,*B*). Finally, streamline loss occurred at an earlier age in females than in males.

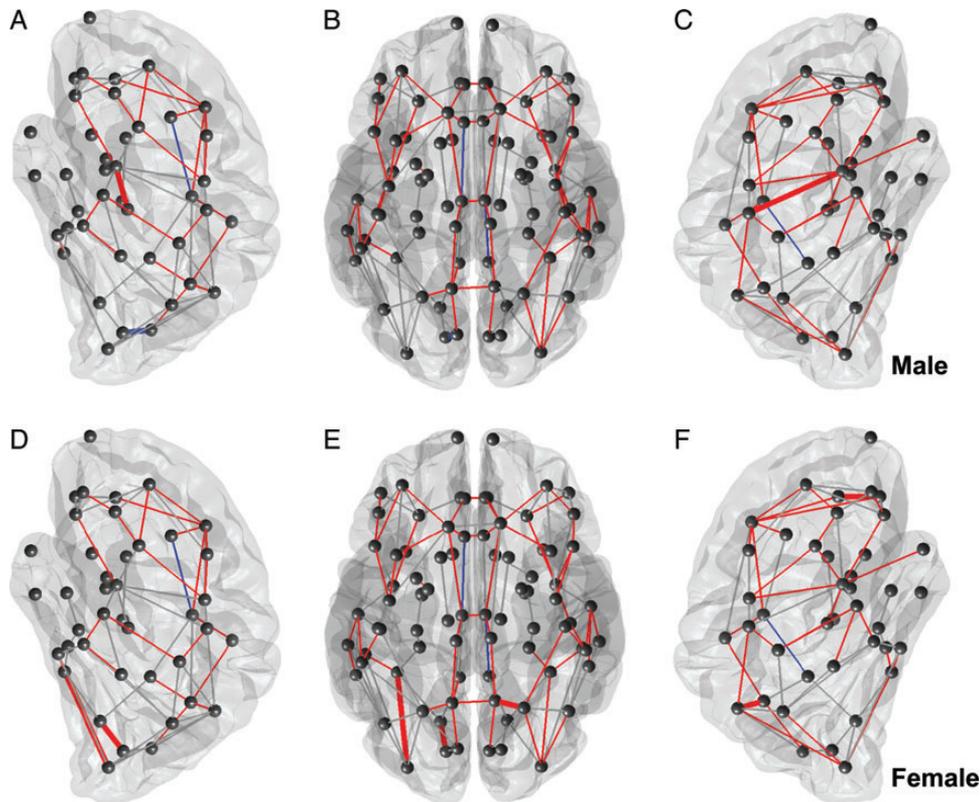

**Figure 4.** Sex-specific developmental changes in individual edge analysis for male (*A–C*) and for female subjects (*D–F*), where red edges represent significant decrease, blue edges indicate significant increases over development, gray edges illustrate the tested edges that all subjects shared in common and the sex-specific changes were emphasized by the thick edges. (*A* and *C*) Sagittal views of the left hemisphere, (*B* and *D*) transverse view, and (*C* and *F*) sagittal views of the right hemisphere, of male and female brains, respectively. (*A*) Two edges showed age-related changes; one in the temporal lobe lost streamlines and the other edge in the occipital lobe gained streamlines. (*C*) An edge in the parietal lobe lost streamlines. (*D*). Two edges in the temporal and the occipital lobes lost streamlines. (*F*) Two edges in the frontal and parietal lobes lost streamlines.

### Stable Small-World and Modular Organization with Preferential Streamline Loss Within Short-Distance, Thick, and Intramodular Fiber Tracts

We found that fewer long-distance, thin, and intermodular fiber tracts showed streamline loss than would be expected given how often such fiber tracts could have been affected by chance. This preferential streamline loss has several implications for the stable topological features that we observed. First, we found that small-world features were retained over age despite the overall reduction in the number of streamlines. A significant decrease in many long-distance streamlines would remove shortcuts and result in larger path lengths and reduced global efficiency while fewer connections between neighbors would decrease local clustering and local efficiency, disrupting small-world features of a brain network. However, global efficiency stayed comparable with that of rewired networks, local efficiency was much higher than in rewired networks across age, conserving small-world topology (Latora and Marchiori 2001, 2003). We would therefore expect changes mainly in short-distance connectivity. Indeed, short streamlines were mostly affected and long-distance connectivity was rather preserved. Relatively conserved streamlines in long-distance fiber tracts could be achieved by strengthening long-range pathways while a reduction in the number of streamlines in short fiber tracts could be due to weakening of short connections, which is consistent with previous findings from rs-fMRI and DTI data (Fair et al. 2009; Supekar et al. 2009; Dosenbach et al. 2010; Hagmann et al. 2010).

Second, in line with previous rs-fMRI and DTI studies (Fair et al. 2009; Hagmann et al. 2010), modularity *Q* remained stable over age. We found that the global modular organization and module membership of ROIs were unchanged with local changes especially in the basal ganglia. Therefore, local networks re-organized their relationships with other community members while keeping the global community structure stable. This retained modular organization (Kaiser et al. 2010; Meunier et al. 2010) might be crucial in keeping the balance between information integration and the segregation of separate processing streams (Sporns 2011). Too many connections between modules would interfere with different processing demands, for example, leading to interference between visual and auditory processing. In addition, more intermodule connections would also facilitate activity spreading potentially leading to large-scale activation as observed during epileptic seizures (Kaiser et al. 2007; Kaiser and Hilgetag 2010).

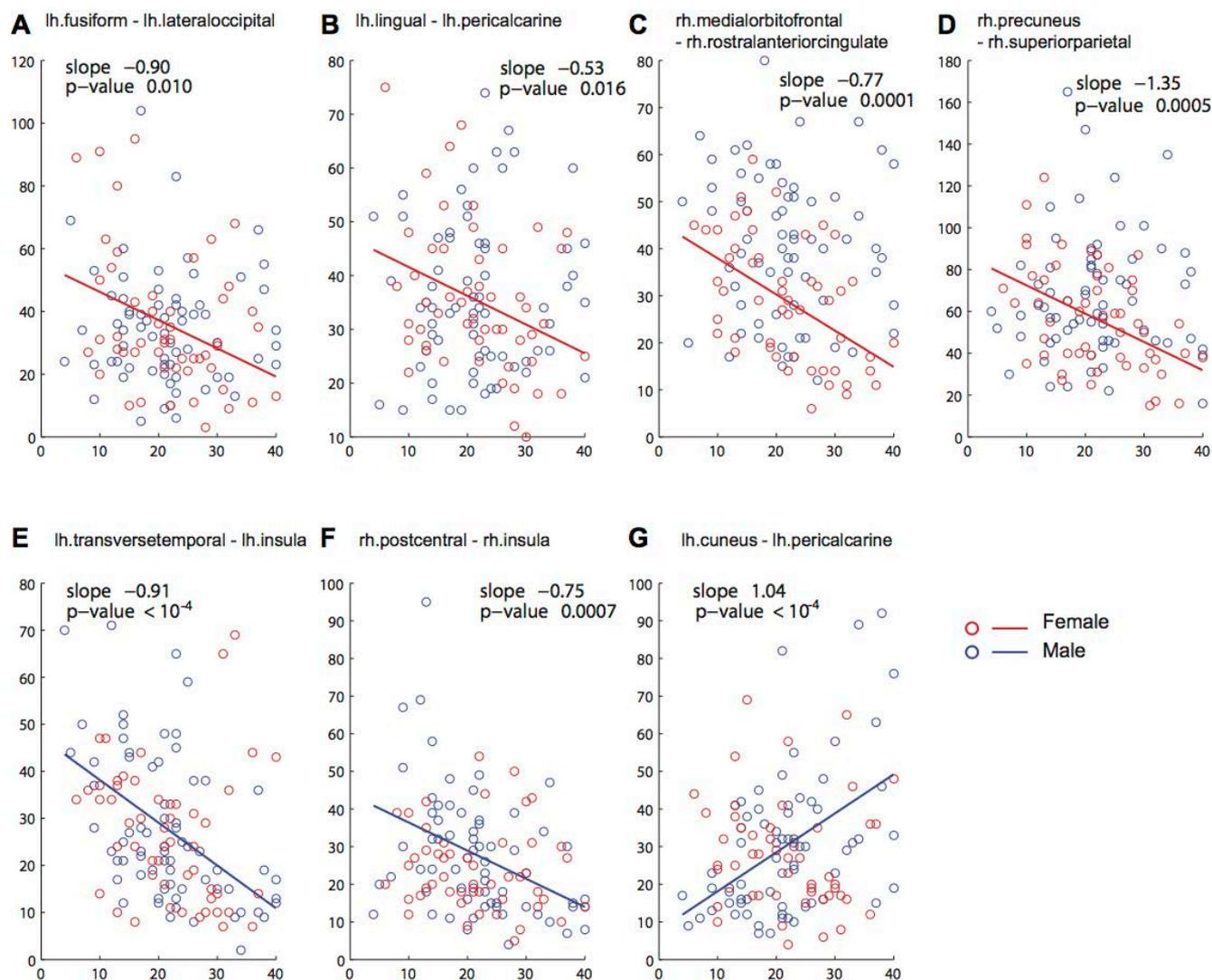

**Figure 5.** Sex-specific developmental changes. (*A–G*) Scatter plots of streamline count with relevant fitted lines. Red: female, blue: male. Upper panel: The 4 fiber tracts demonstrating age effects only for females. Lower panel: The 3 fiber tracts displaying age effects only for males. Lh, left hemisphere; rh, right hemisphere. (*A*) The fiber tract between lh.fusiform and lh.lateraloccipital showing a reduction of streamline counts only for females. (*B*). The fiber tract between lh.lingual and lh.pericalcarine with a decreased number of streamlines for females, (*C*). rh.medialorbitofrontal–rh.rostralanteriorcingulate, (*D*) rh.precuneus-rh.superiorparietal, (*E*) The fiber tract between lh. transversetemporal and lh.insula with a reduced number of streamlines over age only for males, (*F*) rh.postcentral–rh.insula, (*G*) lh.cuneus–lh.pericalcarine. The rate of change per year and corresponding *P* value is included in the figure and FDR-adjusted *P* values can be found in Table 2.

However, because of the reduction of streamlines in intramodule edges, proportionally intermodule connections increased, indicating that the brain network became more distributed rather than modular with age as observed in previous studies, which was associated with development of advanced cognitive abilities by enhancing integration of neural processing (Fair et al. 2009; Supekar et al. 2009; Hagmann et al. 2010).

In summary, we find that long-distance and intermodular connectivity is largely spared from the ongoing streamline losses during development, which is potentially beneficial for the observed stability of small-world and modular connectome features. Note that as connections between modules are not necessarily long distance (Kaiser and Hilgetag 2006), we found that only 47% of intermodular fiber tracts also belong to the class of long-distance connections. Retaining long-distance and intermodular fibers indicate that small-world features, such as the number of processing steps but also the balance between information integration and large-scale brain activity, are kept within a critical range during development (Kaiser and Hilgetag 2006). Preserving this balance is crucial as changes in long-distance connectivity are a hallmark of neurodegenerative and neurodevelopmental disorders ranging from Alzheimer's disease (Ponten et al. 2007; Stam et al. 2007) to schizophrenia (Alexander-Bloch, Vértes, et al. 2012). Therefore, stable topological network features might help to prevent cognitive deficits in neuropsychiatric disorders.

Another important implication of the reduced number of streamlines is the relationship to the number of edges within a network. Changes in streamline count can lead to a reduction of connections within a network if an edge comprised of few streamlines loses all its streamlines, thus reducing edge density. However, edge density did not significantly change during brain maturation. Therefore, several mechanisms are conceivable how the number of edges is maintained during development. One option is that newly emerging edges cancel out disappearing edges (equilibrium state), which is biologically costly by removing already established connections and unlikely because new connections are established mostly early in the development. Alternatively, only the weight of an edge changes (stable state). For the latter case, a reduction of streamlines in thin edges, which could result in the loss of the whole edge, needs to be prohibited. Indeed, we found that thick edges were mostly affected from the decreased streamlines, thus preserving the structure of the network. This is beneficial, as reducing thin fibers would necessitate an increase in synaptic weights or number of synapses to transmit the same amount of information. Reducing streamlines for thick fibers, on the other hand, has only a small effect on activity flow due to the large number of remaining streamlines.

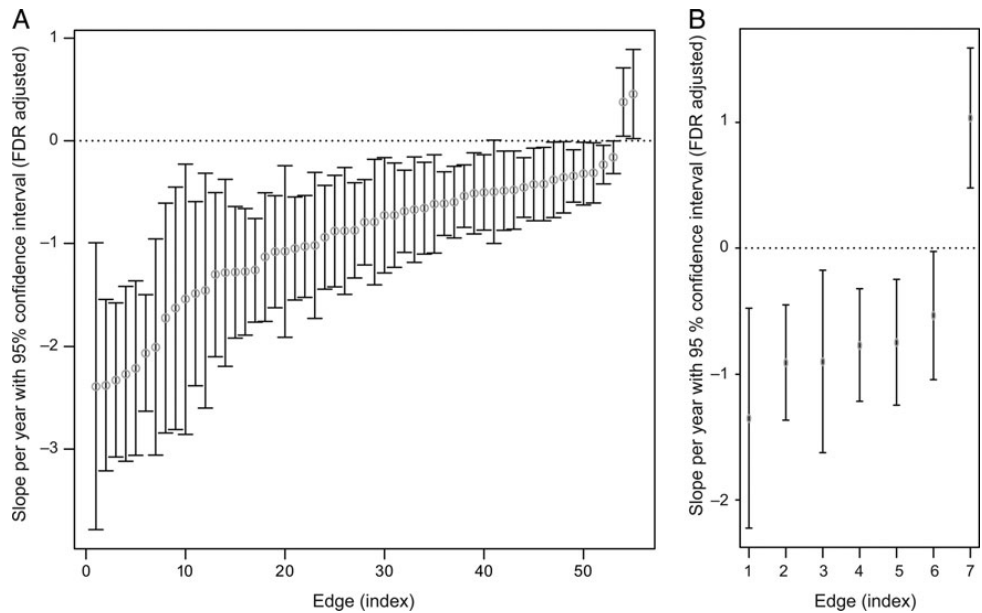

**Figure 6.** Individual edge slopes representing age effect per year with FDR-adjusted confidence intervals. (*A*) Individual edge age effect for both genders. *x*-axis: indices of edges, *y*-axis: coefficients for age effect per year with FDR-adjusted confidence intervals. The last 2 edges with positive slopes and confidence interval ranges are the edges with an increased streamline count and the others are the fiber tracts characterized by a decreased number of streamlines. (*B*) Age-related sex effect. *x*-axis: indices of edges, First 4 edges show decreasing rate of streamline count for females and the rest 3 edges displays age effect for males, *y*-axis: coefficients for age effect per year with FDR-adjusted confidence intervals.

### Preferential Streamline Loss for Frontal and Subcortical Regions

Changes in individual edges were most pronounced in the frontal lobe, a brain region that is characterized by protracted development until the third and fourth decade of life as indicated by ongoing synaptic pruning and myelination (Benes et al. 1994; Sowell et al. 1999; Shaw et al. 2008; Petanjek et al. 2011). In addition, the fiber tract between putamen and pallidum in the basal ganglia for the left hemisphere was characterized by a reduced number of streamlines. Previous studies that examined GM volume (Sowell et al. 1999) also found changes in GM density in putamen and pallidum in postadolescent brain development, which are involved in learning and neurodevelopment diseases (DeLong et al. 1984; Alexander and Crutcher 1990; Hokama et al. 1995; Teicher et al. 2000; Ell et al. 2006; DeLong Mr 2007; de Jong et al. 2008; Farid and Mahadun 2009). Furthermore, basal ganglia were characterized by decreased within-module strengths and increased participation coefficients over age. This suggests that connectivity to within these areas decreased relative to connections to outside of the basal ganglia, which is consistent with data from Supekar et al. (2009) who demonstrated that subcortical functional connectivity in children had higher degree and efficiency than in adults.

This reorganization of corticosubcortical connectivity could be involved in the ongoing changes of cognition and behavior during development. The basal ganglia involve regions that are crucially involved in neural circuits relevant for response inhibition and reward modulation. Previous studies have shown that response inhibition improves significantly with age (Williams et al. 1999) as well as reward modulation (Gardner and Steinberg 2005). Unlike for the basal ganglia, the ACC was characterized by an increased connectivity within its module with age. This observation is consistent with functional connectivity of ACC that develops a more focal organization with age (Kelly et al. 2009). ACC has also shown to mature late through error-related ERPs (Santesso and Segalowitz 2008).

### Delayed Streamline Loss for Males

Individual edge analysis revealed sex-specific age effects in the occipital and parietal lobe but to a much lesser extent in the frontal lobe. This is consistent with a previous WM study where mainly the occipital lobe development varied with sex while the growth trajectory in the frontal lobe was similar for both genders (Baron-Cohen et al. 2005; Lenroot et al. 2007; Giedd 2008; Perrin et al. 2009). These results can be explained if we assume that the same mechanism of preferential streamline loss operates at different time-scales in males and females. Provided that males had a similar developmental curve but with a shifted peak (Fig. 7*C,D*), we can explain the sex-specific changes. As expected from the shifted peak hypothesis (Fig. 7*C,D*), the total number of streamlines for males, but not females, remained stable at an earlier age range (4–28 years, not shown) while both genders showed streamline reductions in the age range 4–40 years. This delayed developmental growth curve in streamline count can be related to later volume growth peaks for GM and WM in males (Giedd et al. 1997; Giedd and Rapoport 2010) and earlier myelination for females (Benes et al. 1994).

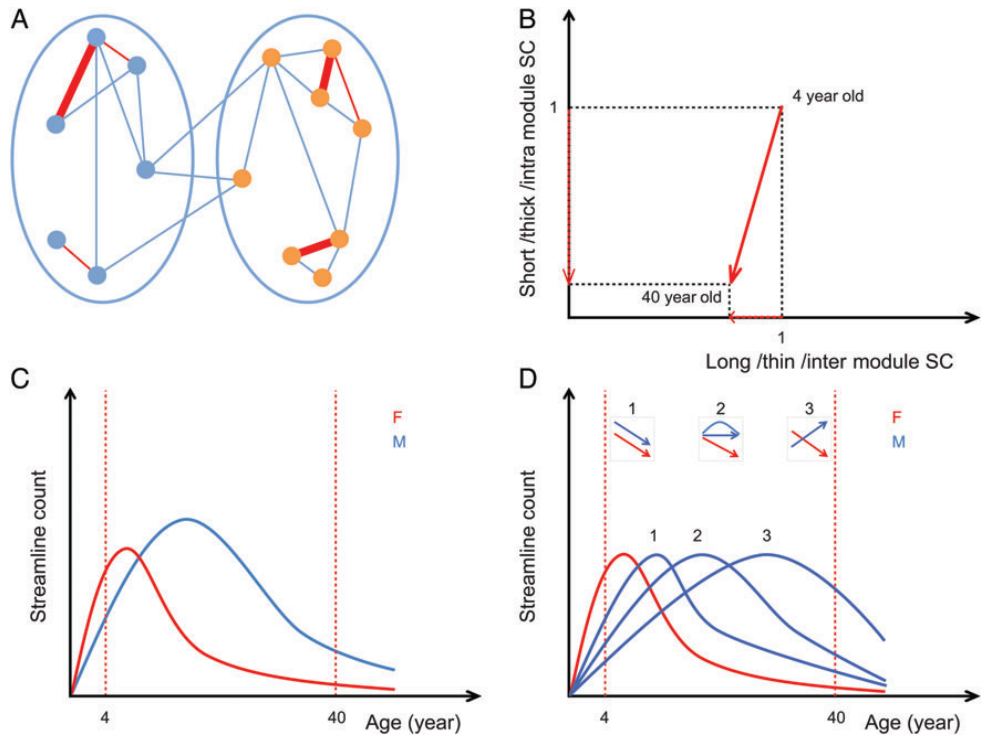

**Figure 7.** (*A* and *B*) The schematic summary of the preferential reduction of thick, short, and within-module streamlines over age. (*A*) Location of change: 2 ellipses represent left and right hemispheres and small circles inside hemispheres indicate ROIs. Lines connecting ROIs illustrate fiber tracts between ROIs. Red lines are where the reduction of streamlines occurred; thick, short or intramodule edges were mostly affected. (*B*) Magnitude of change: Short, thick, or intramodule edges lost more streamlines than long, thin, or intermodule edges. *x*-axis: either long, thin, or intermodule streamline count (SC), *y*-axis: either short, thick, or intramodule SC. (*C* and *D*) Hypothetical developmental curves for males (blue) and females (red). (*C*) For the total streamline count based on the observation of our data (Fig. 2*A*): a longer lasting and higher peaked increase and a delayed decrease in males. (*D*) For individual edges: we observed sex-specific development (Fig. 4*C*), which can be explained by 3 representative cases: if the 2 curves strongly overlap they show similar decreasing patterns (case 1), if one of the curves peaks later, one curve shows a decreasing pattern while the other curve is still increasing (case 3) or simply not decreasing yet (case 2). Therefore, depending on the time scale of the developmental trajectory, males and females may show different patterns.

We only observed circumscribed sex-differences independent of age. Local efficiency was higher for females than males consistent with Gong and colleagues' finding (Gong et al. 2009) and some ROIs showing higher within-module strength and lower participation coefficient for females can be related to higher local efficiency in females. Interestingly, absolute difference in the number of streamlines between genders was not uniformly distributed; males exhibited more streamlines for intramodule edges. This is consistent with the finding that males and females do not differ in the WM volume growth trajectory in the corpus callosum (Giedd 2008). However, this means proportionally females have more connections across hemispheres and between modules (DeLacoste-Utamsing and Holloway 1982; Davatzikos and Resnick 2002; Allen et al. 2003).

### Structural Correlates of Streamline Loss

The observed reduction in the total number of streamlines could be related to rs-fMRI developmental "system-level pruning" (Supekar et al. 2009), considering tight coupling between SC and functional connectivity (Honey et al. 2009, 2010). As Supekar and colleagues suggested for functional connectivity (Supekar et al. 2009), the decreased number of streamlines for short and intramodule connections in this study could be due to weakening of local connections through synaptic pruning and neuronal rewiring. These local processes prolong until adulthood and are major factors for anatomical developmental changes (Benes et al. 1994; Petanjek et al. 2011). The reduction of synapses and corresponding axons or axon collaterals could potentially also lead to a decreased number of streamlines within fiber tracts. Owing to technical limitations of DTI, pruning of dendrites and intracortical connections cannot be detected. However, synaptic pruning in the prefrontal cortex for intracortical connections (Petanjek et al. 2011) was mainly limited to children at younger ages than in our study (Petanjek et al. 2008). In contrast, the pruning of long-distance connection, observable in DTI, occurs in developing rhesus monkeys, both at earlier and later stages of development (LaMantia and Rakic 1990, 1994; Luo and O'Leary 2005). Considering both limitations of DTI (Jones and Leemans 2011) and previous studies (Fair et al. 2009; Supekar et al. 2009; Dosenbach et al. 2010), changes in corticocortical and subcorticocortical projections might underlie our results but further investigations are needed to determine the contributions of these potential biological correlates.

Studies have shown that volume for WM fiber tracts increased with age (Faria et al. 2010; Lebel and Beaulieu 2011) and continued myelination also leads to an increase in WM volume, which could explain an increase in total WM volume while undergoing a possible reduction of fiber tracts. Even though streamlines were reduced in our study, an increased myelination might still have taken place but might have been overshadowed by axonal changes and vice versa. Greater amounts of myelination would generate higher FA values (Mädler et al. 2008; Faria et al. 2010), leading to an increase in the number of detected streamlines. For example, even if the number of axonal projections were reduced the remaining fibers with an increased myelination could be detected easily by tractography and compensate the lost fiber tracts, leading to no changes in the number of streamlines. Thus, the balance between myelination and axonal pruning may have contributed to our final results.

The reduction in streamlines with age cannot be attributable to ongoing changes in the number of seed voxels used for tractography as this number was unaffected by age. Other factors affecting tractography include axon diameter distributions (See detailed Discussion Jones 2010; Jones et al. 2013) and fiber curvature changes. If many fiber tracts became more curved over age, as DTI normally does not track highly curved trajectories, the number of streamlines of the fiber tract could decrease. However, most of the fiber tracts (edges) that we tested did not change their curvature over age (83%, 106 of 128) only 22 edges (17%) showed changed curvature over development. Of these 22, only half showed curvature increase. For a single edge we find streamline decrease while curvature increased ruling out curvature as a confounding factor of our results (See detailed Results and Discussion in Supplementary Material S4).

### Limitations

Even though the current study observes a large dataset, there are several inherent limitations. First, the subjects were unequally distributed across ages. Having subjects at ages between 4 and 40 years may not be optimal for detecting major changes as small-world and modular features were established during the first 2 years (Fan et al. 2011; Yap et al. 2011). Our focus, however, was not the major structural changes but the continuous development while keeping the network economic (Vertes et al. 2012) and stable. Second, studies that network approaches use different definitions for weight and different normalization schemes complicating the comparison between studies. We used absolute number of streamlines as weights; however, our results are consistent with previous studies with slightly different weight definitions (Gong et al. 2009; Hagmann et al. 2010). Third, our DTI approach, unlike DSI or HARDI analysis, will not resolve crossing fibers. However, the shorter recording time of this data are an advantage when measuring connectivity in children. Modeling through probabilistic tracking with crossing fibers (Behrens et al. 2007; Jbabdi and Johansen-Berg 2011) would therefore be a future research direction. Although streamlines do not directly correspond to axonal projections (Jones 2010; Jones et al. 2013), we found our results were consistent with previous anatomical studies (Benes et al. 1994; Sowell et al. 1999; Gong et al. 2009; Perrin et al. 2009).

### Conclusions

The human brain undergoes vast structural changes during development. Nonetheless, brain networks develop in a way that preserves its topological (small-world/modular) and spatial (long-distance connectivity) organization to secure its capability of integration of information and individual processing of modules. This present study showed how brain connectivity changed during development in terms of fiber tracts as well as global network features. We showed preferential decreases in the number of streamlines for thick, short-distance, and within-module/within-hemisphere fiber tracts. These changes may not necessarily occur at the same time for males and females; males seem to show a delayed start from the prolonged development in WM and GM. However, although with different time courses between genders, the global topological features ensuring healthy brain development apply to both genders. Therefore, brain networks maintain their topological stability during brain development by preferentially modifying structural connectivity.

### Supplementary Material

Supplementary material can be found at: http://www.cercor.oxfordjournals.org/.

### Funding

This work was supported by National Research Foundation of Korea funded by the Ministry of Education, Science and Technology (R32-10142 to S.L., C.E.H., and M.K.), the National Research Foundation of the Korea government (MSIP NRF, 2010-0028631 to C.E.H.), the Royal Society (RG/2006/R2 to M.K.), the CARMEN e-Science project (http://www.carmen.org.uk) funded by EPSRC (EP/E002331/1, EP/K026992/1, EP/G03950X/1 to M.K.), and Max-Planck Society to P.J.U. Funding to pay the Open Access publication charges for this article was provided by EPSRC. *Conflict of Interest*: None declared.